# Behavior Revealed in Mobile Phone Usage Predicts Credit Repayment

Daniel Björkegren and Darrell Grissen

**ABSTRACT**

Many households in developing countries lack formal financial histories, making it difficult for firms to extend credit, and for potential borrowers to receive it. However, many of these households have mobile phones, which generate rich data about behavior. This article shows that behavioral signatures in mobile phone data predict default, using call records matched to repayment outcomes for credit extended by a South American telecom. On a sample of individuals with (thin) financial histories, our method actually outperforms models using credit bureau information, both within time and when tested on a different time period. But our method also attains similar performance on those without financial histories, who cannot be scored using traditional methods. Individuals in the highest quintile of risk by our measure are 2.8 times more likely to default than those in the lowest quintile. The method forms the basis for new forms of credit that reach the unbanked.

**KEYWORDS**: credit scoring, machine learning, digital credit, mobile phones, financial inclusion

**JEL CODES**: D14, O16, G23, O12, C58

This draft: January 28, 2019. First draft: January 6, 2015.

**ACKNOWLEDGEMENTS**

Daniel Björkegren (corresponding author) is a professor at Brown University, Providence, Rhode Island, USA; his email address is dan@bjorkegren.com. Darrell Grissen is a former employee of the Entrepreneurial Finance Lab, and was employed during data collection; his email address is dgrissen@gmail.com. The research for this article was supported financially by the George Shultz Fellowship from the Stanford Institute for Economic Policy Research and the W. Glenn Campbell and Rita Ricardo-Campbell National Fellowship at Stanford University. The authors are grateful to Entrepreneurial Finance Lab and partners for providing data. The authors thank Jeff Berens, Nathan Eagle, Alfredo Ebentreich, Javier Frassetto, and Seema Jayachandran for helpful discussions, as well as audiences at the American Economic Association Annual Meetings, NEUDC, Microsoft Research, NBER IT and Digitization, and the AMID/BREAD Summer School in Development Economics.

# 1. INTRODUCTION

Mobile phones have spread dramatically: there are over 4.5 billion mobile phones in developing countries (ITU, 2011). In addition to improving communication, these devices make it possible to provide new forms of credit. Telecoms often extend credit to help smooth phone consumption itself: as incomes rise, consumers are expected to transition from prepaid to postpaid plans, which are essentially rolling credit accounts. Phone networks can also be used to mediate loans for consumer durables, such as pay-as-you-go solar energy systems.[1] And through mobile money, phone networks can provide loans directly to consumers at close to zero transaction costs. This can enable forms of credit different from microcredit or traditional bank lending (consider M-Shwari in Kenya, or independent smartphone lending apps).

However, all of these applications face a fundamental problem: how can a lender assess whether a potential borrower will repay? Few developing country residents have traditional credit scores. Few interact with formal institutions at all: 2 billion people lack bank accounts (Demirguc-Kunt, Klapper, Singer, & Van Oudheusden, 2014). And in person methods, like interviews or peer groups in microfinance, are costly for small loans and remote households.[2]

This paper develops a method to predict repayment of the poor at scale. Its key insight is that most poor consumers do have at least one rich record of formal interaction—their interaction with a mobile phone itself. The way that a person uses a mobile phone can predict whether they will repay credit.

This project develops and assesses a low cost method to predict repayment of credit using mobile phone metadata, which are already being collected by mobile phones. From raw phone transaction records it extracts signals plausibly related to repayment, and uses a machine learning approach to combine these signals into a prediction of repayment. Using only minor tweaks to off-the-shelf machine learning methods, it demonstrates performance comparable to that of traditional credit bureau models, including for unbanked individuals who cannot be scored by traditional methods.

---

[1] Typically, if approved for a household solar energy loan, consumers receive the device and pay installments via mobile money. If consumers miss payments, their device is remotely switched off via the mobile phone network.

[2] Nonpayment is particularly problematic in developing societies, as creditors have little recourse if a borrower were to default: borrowers have little in the way of collateral, and systems for legal enforcement are limited.

Since the online posting of this paper's proposal (Björkegren, 2010) and working paper (in 2015), our approach has received substantial attention (NPR, 2015), and is already widely implemented. Combined with mobile money, it forms an essential component of a new ecosystem of digital financial services. The first digital credit product (M-Shwari) was launched in Kenya in 2012, evaluating risk using a simple scoring rule based on mobile money usage. In a few short years, it has been followed by over 68 other digital credit products, which use risk scoring approaches similar to ours. Digital lending smartphone apps have attracted over $190m in venture capital funding. So far, most of these loans are small, but they are already widespread: in Kenya alone, over $2b has been disbursed, and 27% of the population has taken out a digital loan (Totolo, 2018). More Kenyans have loans through these new digital platforms than through traditional banking, or microfinance (FSD, 2016).

Our method consumes raw metadata from mobile phone usage, which are already being collected at close to zero cost. These records can yield rich information about individuals, including mobility, consumption, and social networks (Blumenstock, Cadamuro, & On, 2015; Gonzalez, Hidalgo, & Barabasi, 2008; Lu, Wetter, Bharti, Tatem, & Bengtsson, 2013; Onnela et al., 2007; Palla, Barabási, & Vicsek, 2007; Soto, Frias-Martinez, Virseda, & Frias-Martinez, 2011). This paper shows how indicators derived from this data can predict the repayment of credit.

There are many straightforward indicators of behavior that are plausibly related to repayment of credit. For example, a responsible borrower may carefully manage their balance over time so usage is more smooth. An individual whose usage repeats on a monthly cycle may be more likely to have a salaried income. Or, an individual whose calls to others are returned may have stronger social connections that allow them to better follow through on entrepreneurial opportunities.

From raw transaction records, we extract approximately 5,500 behavioral indicators. Determining how to extract relevant behaviors from unstructured transaction records is the critical step in standard machine learning ('applied machine learning is basically feature engineering' (Ng, 2011)). A brute force data mining approach such as Blumenstock et al. (2015) would algorithmically extract indicators while being agnostic towards the outcome variable. However, such approaches can pick up spurious correlations that make them unreliable in practice (Lazer, Kennedy, King, & Vespignani, 2014). With infinite data and variation, a machine learning approach would drop spurious variables on its own; but when data and variation are finite, focusing on features with intuitive or theoretical links to the outcome of interest can improve stability. Intuition can also suggest more nuanced indicators which a brute force method would neglect. Our approach generates indicators intuitively linked to repayment, which may pick up how consumers manage usage over time, social connections, and consumers' potential capacity to repay. Indicators that have an intuitive link are also more palatable to implementation partners who can be wary of 'black box' methods.

This paper show that these behavioral indicators can reduce the uncertainty around a person's type. They can derisk interactions between large firms and the poor, at scale. The largest promise of these new technologies is not to duplicate the familiar loan offerings that evolved under the constraints of physical interaction, but to enable new types of formal lending that would not be feasible under historical constraints.

This paper demonstrates the method with data from a telecom in a middle income South American country that is transitioning subscribers from prepaid to postpaid plans. In this country, only 34% of adults have bank accounts but 89% of households have mobile phones.

It tends to be difficult to quantify the performance of scoring rules in mature applications because of a 'missing labels' problem: one only observes repayment among people who were granted credit in the first place. As a result, if a lender already uses a sophisticated selection rule that fully incorporates all available information, little can be learned about the performance of different rules. In this regard, our setting has two crucial features. First, in the exploratory phase observed, the telecom sought to learn the risk of different types of applicants, and so extended credit permissively. It selected subscribers with sufficient usage from across the distribution of credit histories (including no history at all). Our results are thus directly applicable to subscribers with a sufficient level of mobile phone usage (performance is similar with lighter usage). Within this population, the data include outcomes from a distribution of individuals who the telecom might conceivably wish to transition to a postpaid plan. As a result, we can evaluate the performance of any screening rule, including ones that use credit bureau information, phone indicators, or both. Second, our sample includes both banked and unbanked consumers, which allows us to both benchmark our performance against credit bureau models, and also evaluate whether performance is comparable for individuals without bureau records. The data include each applicant's mobile phone transaction history prior to the extension of credit, and whether the credit was repaid on time. We predict who among these individuals ended up repaying, based on how they used their mobile phones prior to the switch, in a retrospective analysis. Our data include call and SMS metadata, but not mobile money or top up information. Performance is expected to improve with richer data and larger samples, observed over longer time periods.

After developing our method, this paper presents three main findings.

First, the method has the potential to achieve useful predictive accuracy, even with standard machine learning models. Performance is assessed in two steps. First, in a sample of formally banked (but thin file) consumers, the method actually outperforms credit bureau models, which perform relatively poorly (we consider the industry standard measure Area Under the ROC Curve or AUC, which ranges from 0.61-0.76 for our models versus 0.51-0.57 for bureau models). Second, the method performs similarly well for unbanked consumers, who cannot be scored with traditional methods (AUC 0.63-0.77). Our models perform within the (wide) range of published estimates of traditional credit scoring in the literature (AUC 0.50-0.79). Individuals in the highest quintile of risk by our most conservative measure are 2.8 times more likely to default than those in the lowest quintile. Among those with credit histories, if credit were extended to the 50% lowest risk prospects according to the credit bureau the default rate would be 9.7%, whereas it would be only 8.3% based on our scoring using phone records. Moreover, if credit were extended to those without credit histories whose predicted risk of default would place them in the top 50% of risk-prospects for those with credit records, the default rate would be only 6.6%. Our method can identify a group of good credit prospects from among those with no credit history.

Second, care must be taken to ensure stability over time. In practice, a creditor would use past performance to train the model that disperses future credit. The performance of machine learning methods can deteriorate if the underlying environment shifts over time (Butler, 2013; Lazer et al., 2014). The most straightforward way to set up the prediction task can pick up coincident shocks in addition to underlying factors correlated with repayment. This paper develops a technique to minimize this form of intertemporal instability, by using only variation within each time period to differentiate individuals (analogous to a form of temporal fixed effects). This technique improves intertemporal stability; our models continue to outperform bureau models in our setting when estimated and tested on different time periods.

Third, information currently in the credit bureau is only slightly complementary to that in our indicators. This suggests that in contexts with thin bureau files, there may be limited gains from integrating these new forms of credit with the information already in legacy traditional credit bureaus. However, once digital credit is granted using our method, repayment can be reported to bureaus, which can improve incentives to repay as well as the amount of information in the bureau.

The paper most similar to this is Pedro, Proserpio, and Oliver (2015), which finds that among individuals with active credit cards, those who recently defaulted have different calling behavior afterwards than those who repaid successfully, using 58 indicators of behavior. However, a person is likely to alter their patterns of calling after a default. It is unclear whether the basic indicators that paper derives could predict default, or simply pick up shocks that are only correlated with default ex post. Additionally, that paper observes only individuals who were already screened through a traditional credit bureau, so it cannot assess performance relative to a benchmark, or whether phone indicators can be used to screen the unbanked.

A short companion paper uses a simple model to illustrate how improved screening can affect access to credit, and analyzes additional heterogeneity results (Björkegren & Grissen, 2018).

Our approach is one of several nontraditional alternatives that can identify individuals who are likely to repay credit. For example, psychometric tests are also predictive of repayment among both the banked and unbanked (Arráiz, Bruhn, & Stucchi, 2017).[3] These tests are typically designed to directly capture behaviors that have a tighter theoretical link to repayment, such as ability and willingness to repay. However, they require additional interaction with applicants, which imposes costs.

## 2. CONTEXT AND DATA

The primary organizational partner is EFL (Entrepreneurial Finance Lab), which works on alternative credit scoring methods in developing and emerging markets, with an emphasis on the underbanked.[4] EFL identified a partner that was interested in exploring alternate methods of assessing creditworthiness.

As consumers in emerging economies have become wealthier, many telecoms have begun transitioning their subscribers from prepaid plans to postpaid subscriptions. Under postpaid plans, subscribers do not face the hassle of topping up their account, and so tend to consume more, and are also less likely to switch to competitors. However, postpaid plans expose the telecom to the risk that a subscriber runs up a bill that they do not repay. In developed countries, many telecoms check subscribers' credit bureau files before granting a postpaid account. However, in lower income countries these files are often thin, or nonexistent. The authors partnered with a telecom in a middle income South American country, with GDP per capita of approximately $6,000, which sought to transition a subset of its prepaid subscribers to postpaid plans.[5] It wished to expand this subset to include those with sparse or nonexistent formal financial histories.

---

[3] This paper was also written in collaboration with our organizational partner, EFL, which also does psychometric credit scoring.

[4] From their website, "EFL Global develops credit scoring models for un-banked and thin-file consumers and MSMEs, using many types of alternative data such as psychometrics, mobile phones, social media, GIS, and traditional demographic and financial data. We work with lenders across Latin America, Africa and Asia." http://www.eflglobal.com Darrell Grissen was employed by EFL while data was collected for this paper.

[5] All results reported in US dollars.

The telecom offered a preselected set of subscribers the chance to switch to a postpaid plan with lower rates, and recorded who among these subscribers paid their bills on time. Because the telecom wanted to learn about the risks of transitioning different types of users in this initial exploration, it was permissive and unsophisticated in selecting customers to transition. It selected customers who used their phones sufficiently (who were more likely to benefit from postpaid billing) from across the distribution of credit bureau records.[6] These selected subscribers received a call inviting them to transition to a postpaid plan; those who opted in were switched from prepaid to the cheapest postpaid plan (providing approximately $30 in revolving credit each month). The data cover 7,068 subscribers who were offered postpaid plans and opted in, which is the relevant sample for assessing performance for the telecom. The telecom was aware that paying a phone bill was new for these subscribers, so if a subscriber did not pay their postpaid bill, they were notified by SMS and other channels that their bill was soon to become overdue. If consumers were more than 15 days overdue, their service was cancelled and they were reported to the credit bureau.[7] In our sample, 11% of consumers defaulted.[8] While this form of credit has different features from a traditional bank loan, so do many emerging forms of digital credit; for example, short term loan ladders are common: (Carlson, 2017). For each subscriber, the telecom pulled mobile phone transaction records (Call Detail Records, or CDR). In this setting, many subscribers also had formal financial histories maintained at the credit bureau; the telecom also pulled these records. Bureau records include a snapshot of the number of entities reporting, number of negative reports, balances in different accounts (including consumer revolving, consumer nonrevolving, mortgage, corporate, and tax debt), and balances in different states of payment (normal, past due, written off). It also includes the monthly history of debt payment over the past 2 years (no record, all normal, some nonpayment, significant defaults), and includes a summary score that

---

[6] The authors were not able to obtain the telecom's selection rule, or information on those who were selected out. The data is consistent with the telecom making unsophisticated selection in order to assess the risk of different types of applicant. In the resulting sample, 15% are missing credit bureau records, 26% have perfect bureau summary scores, 41% have near-worst bureau summary scores, and the remainder have summary scores in between. (See Figure S2.)

[7] For many consumers this would be the first record in their credit history. After this point, the consumer could use a prepaid account. Because the telecom could pause service, the credit could be thought of as one with the subscriber's phone number held as collateral. However, that collateral is limited, as subscribers could open a new prepaid account with a new phone number.

[8] This rate is higher than typical microcredit loans in this context.

combines these indicators according to the bureau's judgment of what factors are important (using a decision rule not trained on data). Subscribers were matched to their financial histories based on an encrypted, anonymized identifier.

The mobile phone data include metadata for each call and SMS, with identifiers for the other party, time stamps, tower locations, and durations. It does not include top-ups, balances, data access, charges, handset models used, or mobile money transactions; thus performance is expected to improve with richer data. The data do not include any information on the content of any communication.

We aim to predict default based on the information available at the time credit was granted, and so include only mobile phone transactions that precede the date of plan switching. Descriptive statistics for the sample are presented in Table 1. Although 85% of our sample has a file at the credit bureau, many of these files are thin: 59% have at least one entity currently reporting an account, 31% have at least two, and only 16% have at least three. By construction, 100% of the sample has a prepaid mobile phone account. The median individual places 26 calls per week, speaking 32 minutes, and sends 24.4 SMS. The data includes the median individual's phone usage for 16 weeks; an implementation that can obtain longer histories is likely to perform better.

## 3. METHOD

The goal is to predict the likelihood of repayment using behavioral features derived from mobile phone usage. We consider a sample of completed plan transitions, and consider whether information that was available at the time the credit was extended could have predicted its repayment. Because this sample of individuals did obtain credit, risk is reported among those who received credit based on the selection criteria at the time, which was relatively permissive and spanned the distribution of credit histories (including no history at all).

The credit data provide an indicator for whether a particular borrower repaid their obligation (the partner's definition is 15 days past due). From the phone data we derive various features that may be associated with repayment. In a similar exercise, Blumenstock et al. (2015) generates features from mobile phone data using a data mining approach that is agnostic about the outcome variable. Our approach is instead tailored to one outcome, repayment. We extract a set of objects that may have an intuitive relationship to repayment, and then compute features that summarize these objects. We focus on features with an intuitive relationship because implementation partners can be wary of 'black box' methods, and indicators that have a theoretical link are more likely to have a stable relationship to the outcome of interest. While our approach is likely to extract some features similar to Blumenstock et al. (2015), it will also measure more nuanced features that would not have been generated by a generic method.

Phone usage captures many behaviors that have some intuitive link to repayment. A phone account *is* a financial account, and captures a slice of a person's expenditure. Most of our indicators measure patterns in how expenses are managed, such as variation (is usage erratic?), slope (is usage growing or shrinking over time?), and periodicity (what are the temporal patterns of usage?). In particular, individuals with different income streams are likely to have different periodicities in expenditure (formal workers may be paid monthly; vendors may be paid on market days). We also capture nuances of behavior that may be indirectly linked to repayment, including usage on workdays and holidays, and patterns of geographic mobility, which can reflect information on employment. Although social network measures may be predictive (who one is connected to may reflect one's level of responsibility or ability to access resources), we include only basic social network measures that do not rely on the other party's identity (degree, and the distribution of transactions across contacts), as we are hesitant to suggest that a person's lending prospects should be affected by their contacts. While many traditional credit scoring models aim to uncover a person's fixed type (whether the person is generally a responsible borrower), the high frequency behavior we capture may also pick up features specific to the time when an individual is being evaluated for credit (a person may be likely to repay this credit, even if they are not generally responsible). Our process has three steps:

First, the method identifies atomic events observed in the data, each represented as a tuple *(i, t, e, $X_{iet}$)*, where *i* represents an individual, *t* represents the timestamp, *e* represents an event type, and $X_{iet}$ represents a vector of associated characteristics. Event types include transactions (call, SMS, or data use), device switches, and geographic movement (coordinates of current tower). Characteristics derived from the raw transaction data include variables capturing socioeconomics (the handset model, the country of the recipient), timing (time until the credit is granted, day of the week, time of day, whether it was a holiday), and management of expenses (whether the sender or receiver had pre- or post-paid account, whether the transaction occurred during a discount time band, or at the discontinuity of a time band).

Second, for each individual $i$, event type $e$, and characteristic $k$, we compute a vector with the sum of events of each potential value of the characteristic:

$$D_{iek} = \left[ \sum_t 1\{X_{ietk} = d\} \right]_{d \in unique(X_{ek})}$$

This generates, for example, the count of calls by time of day, the number of minutes spoken with each contact, the number of SMS to pre- and post-paid accounts, and the total duration of calls immediately before and after the start of a discount time band.

Finally, for each vector we compute a set of summary statistics. For sequences, these include measures of centrality (mean, median, quantiles), dispersion (standard deviation, interquantile ranges), and for ordinal sequences, change (slope) and periodicity (autocorrelation of various lags, and fundamental frequencies—which correspond to the periods of the strongest repeating temporal patterns. For counts by category, we compute the fraction in each category and overall dispersion (Herfindahl-Hirschman Index). For geographic coordinates, we compute the maximum distance between any two points, the distance from the centroid to several points of interest, and use a clustering algorithm to identify important places (Isaacman et al., 2011). We also compute statistics that summarize pairs of sequences, including correlations, ratios, and lagged correlations (e.g., the correlation of minutes spoken with SMS, which may indicate whether a person coordinates in bursts of activity).

These three steps generate various quantifications of the intuitive features presented (including strength and diversity of contacts) as well as other measures (intensity and distribution of usage over space and time, and mobility). For each feature, we also add an indicator for whether that individual is missing that feature. Altogether, there are approximately 5,500 features with variation.

## 4. PREDICTION AND RESULTS

A first question is how individual features correlate with default. Table 2 presents the single variable correlation with default.

Characteristics traditionally available to lenders are not very predictive. Demographic features (gender and age) have very low correlation with repayment (magnitudes between 0.04 and 0.07). Having a credit bureau record has a small negative correlation with repayment (-0.02). For individuals with records, the most predictive feature is the summary score (-0.072; lower is better), and the fraction of debt lost (-0.046). That individual credit bureau features are only slightly predictive suggests that predicting repayment in this setting is a difficult problem.

Individual features derived from mobile phone usage have slightly higher correlations, ranging up to 0.16. But mobile phone usage data are richer, so there are many more features of behavior that can be included in a model. Many features measure similar concepts, so the table presents broad categories, and the correlation of one top feature within that category. Correlated features include the periodicity of usage (top correlation -0.16), slope of usage (0.13), correlations in usage (0.11), and variance (-0.10). The table highlights particular features that perform well in isolation, including the slope of daily calls sent, and the number of important geographical location clusters where the phone is used. Next, we consider using multiple features together to predict repayment.

**<u>Predicting Repayment</u>**

Our features are predictive even using standard methods common in the machine learning literature. We estimate two standard machine learning models: random forests, and logistic regressions using a model selection procedure (stepwise search using the Bayesian Information Criterion or BIC), for bureau indicators and phone indicators (CDR).[9] Random forests are a generalization of decision trees designed to reduce overfitting, by combining multiple trees which each have access to a subset of the sample (Breiman, 2001).

However, these straightforward estimation routines may muddle the individual factors that explain repayment with common temporal shocks that lead to differences in the proportion of credit repaid in different time periods. High frequency indicators such as our phone indicators are particularly susceptible to picking up these shocks. For phone indicators, we develop two new models that improve intertemporal stability by basing predictions off of only within-week variation (CDR-W). The first is an OLS model with week fixed effects; these absorb week-to-week variation in repayment.[10] Predictions are formed differently from a standard fixed effect model. A standard model would include the fixed effect for each offer week in its predicted repayment, but that is not feasible in this setting: a lender would not know the fixed effect for future weeks. Instead, predictions are based on the average of the past weeks' fixed effects, weighted by the proportion of loans granted in that past week. The second model is an analogous version of random forests: we fit separate random forest models to each past week of data, and combine them in an ensemble. When making a prediction for an individual, each submodel is weighted by the proportion of transitions granted in that past week.[11] This approach reduces the discrepancy between within-time and out-of-time performance; it may also lead to selecting indicators that are more stable over time.

To illustrate the features that the models select, we first estimate these models on the entire sample. For random forest importance plots see Figure S1, and for regression parameter estimates see Table S1, in the supplementary appendix. Standard models tend to place substantial weight on various periodicities of behavior. While some of these patterns are related to repayment, others pick up high frequency artifacts in the data. Our within-week models place less weight on periodicities, and more weight on the fraction of duration spoken during the workday or late at night, the distance traveled, variation in usage, and correlations between calls and SMS. The OLS fixed effect model is also simpler than the logistic model, suggesting the fixed effect approach penalizes model complexity more.

---

[9] The stepwise search is initialized from multiple sets of starting variables; the highest within-fold fit is kept. This article uses the randomForest R package with default tuning of 500 trees, sample sizes of 63.2% drawn for each tree, and $\sqrt{K}$ variables considered at each node (Breiman & Cutler, 2006).

[10] If few transitions are made in a week, it is combined with adjacent weeks.

[11] When giving out loans, one could upweight more recent models to capture changes in conditions.

*Performance*

*Within Time*

We measure how the method performs out of sample using cross validation. Following common practice in supervised machine learning, the sample is divided into $R$ randomly selected folds. The algorithm cycles through each fold, estimating (training) the model on $R-1$ folds, and reporting predictive performance on the $R$th omitted fold (testing). The results are averaged across each fold, and over multiple fold draws. Larger values of $R$ exploit more of the sample for training, which tends to improve predictive performance, but increase the computational burden because the model must be estimated $R$ times. The main results report $R=5$, which is commonly used in the machine learning literature (Table S2 reports results for R=10).

As a first check, consider how well models separate low and high risk borrowers. Results are reported from the most conservative model using our method, the random forest weekly ensemble, and the least conservative bureau model. These models generate continuous scores, so performance can be assessed by comparing a few example acceptance thresholds. In our most conservative model, individuals with the highest quintile of risk scores are 2.8 times more likely to default than those with the lowest quintile. Among those with credit histories, if credit were extended to the 50% lowest risk prospects according to the credit bureau score the default rate would be 9.7%, whereas it would be only 8.3% based on our scoring using phone records. Moreover, if credit were extended to those without credit histories whose predicted risk of default would place them in the top 50% of risk-prospects for those with credit records, the default rate would be only 6.6%. That is, our method using phone records can identify a group of good credit prospects from among those with no credit history.

Because it is not clear where in the distribution of scores a lender would set the acceptance threshold, we trace outcomes along the full range of thresholds. Figure 1 shows how the default rate varies with the fraction of borrowers accepted (where borrowers with lowest predicted default are accepted first).

The receiver operating characteristic curve (ROC) plots the true positive rate of a classifier against the false positive rate, tracing out performance as the acceptance rate is varied. Following the credit scoring literature, we report the area under this curve (AUC) to summarize performance across the range of possible acceptance thresholds.[12] A naïve classifier would generate an AUC of 0.5 and a perfect classifier would generate an AUC of 1.0. Figure 2 illustrates the ROC for the best benchmark model (stepwise logistic) and the most conservative model using indicators derived from phone data (random forest weekly ensemble).

Results for a variety of specifications are presented in Table 3, measuring performance with AUC. (Performance is assessed with alternate metrics in Table S3). Results are presented for the entire sample, and then for the subsamples that do, and do not, have credit bureau records. In this population with thin files, credit bureau information does not perform especially well in predicting repayment (AUC 0.51-0.57). For bureau indicators, the logistic model outperforms the random forest, suggesting that the underlying relationship between those indicators and repayment is relatively linear. In contrast, standard models built on phone indicators (CDR) are predictive, reaching AUCs of 0.71-0.77 when trained and tested on the same time period. Our more conservative CDR-W models achieve lower performance when trained and tested in the same time period (AUCs 0.62-0.63), but also outperform credit bureau models. The performance of our models is also in the range of a sample of published within-time AUC estimates for traditional credit scoring on traditional loans in developed settings (0.50-0.79, shown in Table S4). Our method's performance is similar overall and within each quartile of mobile phone usage, suggesting it picks up nuances in usage rather than overall usage (See Figure S2 and Björkegren and Grissen (2018)). Combining our indicators with information from the credit bureau slightly boosts performance, suggesting that the information gathered by the bureau is only slightly complementary to that collected by our approach.

---

[12] Credit approval decisions tend to approve applicants with scores above a threshold. AUC has two useful properties for these types of decisions: they consider only the relative ranking of observations, and they trace through the range of potential thresholds.

A robustness check assesses models trained instead with 10 fold cross validation in Table S2, which exploits more of the sample but is more computationally demanding. This improves performance of the CDR models in particular, which are more complex and thus data hungry. Their performance is also likely to improve with additional tuning.

Our method also performs better than credit bureau models when assessed with alternate metrics of performance that are informative for our decision problem; in particular, the H-measure designed to overcome some weaknesses of AUC (Hand, 2009) and out of sample $R^2$ (see Table S3). ROC curves are also presented in Figure S4 and comparisons of scores to actual repayment in Figure S5, for the main models.

*Out of Time*

When implemented, a model trained on past data will be used to predict future repayment. As a robustness check, we assess the out-of-time performance of all models by training and testing on different time periods. To do this, we construct an offset version of the dataset. The sample of individuals is split into two; the early group that was transitioned before the median date, and the late group after the median. Then, the phone data is evenly divided, into an early and late period. We construct offset versions of our indicators using only transactions occurring in that half of the data (up to the date of each transition). Because these offset indicators are constructed on a shorter panel, they capture less information than our full indicators. The model is trained on the early group, with phone indicators derived from the early period of phone data, and test it on the late group, with indicators derived from the late period of phone data, with results in the last column of Table 3. Because there is only one late period to test on, out of time results are exposed to much more noise than the within time results (for which performance can be assessed across many fold draws). As a result, these out of time results should be viewed as only a rough check of how much one should trust the within-time results.

Models based on the bureau data, which is lower dimensional, tend to be relatively stable over time. On the other hand, standard models using phone indicators see substantial deterioration (AUC declines from 0.71 to 0.63 for Random Forest and from 0.76 to 0.60 for logistic stepwise). Our modified phone indicator models that use only within week variation are much more stable (AUC increases from 0.62 to 0.64 for Random Forest and decreases from 0.63 to 0.59 for stepwise OLS FE).[13] All phone indicator models continue to outperform models using credit bureau data on this cut of the data (AUC 0.55-0.58). Our performance also lies within the range of the one comparable published benchmark of out of time performance of traditional credit scoring we could find in the literature, from a developed setting (AUC 0.57-0.76, Table S4). Those and our results suggest that bureau models can face at least slight deterioration when tested out of time, with the caveat that we have tested only one time period. The out of time performance of our methods is expected to improve to improve when trained on multiple cohorts (just as credit bureaus have evolved the data they collect by observing default patterns over many cohorts).

After the results from this pilot, the telecom implemented a scoring system using data and methods similar to what we suggested, suggesting they viewed it to be profitable.

## 5. DISCUSSION

Mobile phone data appear to quantify nuanced aspects of behavior that are typically considered soft, making these behaviors 'hard' and legible to formal institutions (Berger & Udell, 2006). Further, these data are already being captured. The method can assist with the provision of financial products to the poor in several ways.

---

[13] The CDR-W Random Forest model is likely to underperform when trained on the same time period with cross validation: it learns less structure when an equivalent sample size is split across multiple time periods (as is the case with out of sample test, which trains on a random subset of loans across weeks).

*Expanding lending to the unbanked*

This paper studies individuals who are near the existing financial system. We summarize the performance of our method by level of formalization in Figure 3. The performance of credit bureau models deteriorates as one moves from individuals with rich financial histories (3 or more entities contributing reports to the bureau) to those with sparser histories. Our method does not deteriorate across levels of formalization, and generates scores of similar performance among individuals with no bureau history, who cannot be scored with traditional methods.

While our method does not require observing a traditional bureau history, it does require observing phone usage. Our partner selected users who spend more on the telephone than the average phone user in the country. We assess selection in Figure S2. We also assess the extent to which performance would deteriorate among sparser users in two exercises, in Björkegren and Grissen (2018), reproduced and updated in that figure. First, we compare performance between quartiles of usage within our sample; our method does not perform worse among users who spend less. Second, we construct a synthetic dataset by dropping transactions from our dataset to match the spending of lighter users. These synthetic datasets simulate the number of transactions that a lighter user might make in the same time span we observe. Performance begins to deteriorate near $1 in spending per month (dropping 96% of transactions). We expect performance for lighter users would improve if they were observed for longer than the median of 16 weeks that we observe. These results suggest that the method may be able to reliably score individuals through most of the distribution of phone usage.

Our approach can dramatically reduce the cost of screening individuals on the margins of the banking system. When poor individuals in the developing world are able to get loans, they are often screened through costly methods such as detailed interviews or peer groups. In contrast, our method can be implemented at extremely low cost, and can be executed over a mobile phone network without the need for physical interaction. These methods enable new forms of lending that do not require the full structure of current branch lending, such as digital credit. Digital credit can both reduce the cost of serving existing markets, and make it profitable to serve consumers outside the current financial system.

*Implementation*

As demonstrated, this approach can be used to extend telecom-specific credit, within the firms that already possess the necessary data.[14] However, the applications are much broader. Mobile money makes it cheap to deliver a loan and collect general payment. With regulatory approval, telecoms may connect to the banking sector, and offer loans to consumers.[15] Alternately, telecoms can package these data into a credit score that can be used by third parties, either through mobile banking platforms or an independent credit bureau.[16] A third implementation, a smartphone app, allows third parties to access usage data independently of telecom operators, and is being explored by several startups.[17] These apps ask for permission to view call history and other behavioral data, and can collect real-time data for a set period.

Our performance estimates are derived from provision of postpaid credit. If phone usage is particularly informative about repayment of this form of credit, performance could differ when predicting default of other forms of loans.[18] But general loans extended over mobile phones have similarities to the postpaid credit we study. Both are extended remotely, without human interaction. In case of nonpayment of a general loan, telecoms can also report borrowers to a credit bureau, and may also be able to freeze a phone account or garnish funds from mobile money balances (as permitted by regulation).[19]

---

[14] In addition to assisting with the transition to postpaid plans, this method can be used to extend credit for handset purcases, or to maintain consistent airtime balances. Many developing country operators offer small airtime loans like this; a scoring model could improve their provision.

[15] See for example, Jumo.

[16] See for example, Cignifi.

[17] See for example, Tala and Branch.

[18] One formulation of this concern would be if heavy phone users are willing to do more to avoid phone account closure, and the method simply picked up the level of phone usage. In that case, it may perform better predicting when repayment of a phone bill than a general loan. However, as mentioned in the text and in Figure S2, the method achieves similar performance in each quartile of usage, and in our sample that has a sufficient amount of usage, the correlation between airtime usage and repayment is very small (-0.03).

[19] Alternately, it could be that bureau information is less predictive for this form of credit than a general loan, which would deflate the benchmark. Bureau information would represent the status quo for the extension of postpaid credit. There is not much evidence on the performance of bureau information in low income populations; the performance observed here is in the lower end of the range of published estimates for general loans from more developed settings (Table S4). However, bureau information

*Privacy*

Privacy will be a key consideration in any implementation. As demonstrated in this paper, the scoring model can be estimated with anonymous data, by anonymizing the identifier that links phone and lending data. However, to generate a prediction for a lending decision, the model must be run on that potential borrower's data. An implementation can be designed to mitigate privacy risks. It can be opt-in, so that only consumers who consent are scored with the system.[20] It can reveal to lenders only a single number summarizing default risk, rather than the underlying features describing behavior. Additionally, it can be restricted to use features that are less sensitive, such as top up behavior rather than the network structure of an individual's contacts.

*Manipulation*

Some indicators are 'gameable' in the sense that a subscriber may be able to manipulate their score if they knew the algorithm. The feasibility of manipulation depends on the complexity of the final model and the susceptibility of individual indicators to manipulation. Both dimensions of the model can be tailored to reduce the probability of manipulation. For example, it is preferable to use indicators that are less susceptible (e.g., manipulating spending or travel can be costly).

*Heterogeneity in performance by subgroup*

An extension to this paper evaluates performance by different subgroups (Björkegren & Grissen, 2018), replicated in Figure S3. The model is trained on all individuals except the omitted fold, and performance is reported for the given subgroup (women, men, and residents in or outside the capital) within the omitted fold. Although our error bars can be wide, performance is not widely heterogeneous across groups, suggesting the method may be able to score different types of individuals.

---

cannot be used to score our population of interest, and our method scores this population at a higher level of performance within this range.

[20] Potential borrowers who opt in may be differentially selected from the broader population, in which case a model estimated on anonymous data from the broader population may not be optimal for use in practice. After the system is operational, it can be periodically refit on outcomes from borrowers who opt in. (Thanks to an anonymous referee for this point).

*If multiple users share each mobile phone account*

In many developing countries, individuals share phones to lower expenses. When a phone account is shared among multiple people, this method will produce one score for the account. The method will still produce an unbiased predictor of the account owner's repayment if sharing practice does not differ between estimation and implementation. In that case, the method will capture both the behavior of phone owners as well as those they choose to share with (indeed the choice of who to share with may also correlate with repayment).

*If each user has multiple mobile accounts*

On the other hand, in competitive mobile markets each individual may use multiple accounts, to take advantage of in-network pricing across multiple networks. This practice is convenient with prepaid plans (with mainly marginal charges) on GSM phones (which allow SIM cards to be easily swapped or may have dual SIM card slots). When users split their call behavior across multiple networks, data gathered from a single operator will represent only a slice of their telephony. While this will make their data sparser, as long as the practice does not differ between estimation and implementation, it will not introduce biases into the method. If individuals use multiple accounts on a single handset (if the handset supports dual SIMs or users swap SIM cards), data gathered from that handset through an app could measure activity across all accounts.

## 6. CONCLUSION

This paper demonstrates a method to predict default among borrowers without formal financial histories, using behavioral patterns revealed by mobile phone usage. Our method is predictive of default in this middle income population, which tends to have thin or nonexistent credit bureau files. In this population which has at least moderate mobile phone spending, our method performs better than credit bureau models. But our method can also score borrowers outside the formal financial system, who cannot be scored with traditional methods. It also performs well with the sparser data that lighter mobile phone users would generate. While this paper is focused on predicting repayment, this type of data can reveal a much wider range of individual characteristics (Blumenstock et al., 2015), and could conceivably be used to predict other outcomes of interest—such as lifetime customer value, or the social impact of a loan.

It has been widely acknowledged that mobile phones can enable low cost money transfers and savings in developing countries (Suri, Jack, & Stoker, 2012). Our results suggest that nuances captured in the use of mobile phones themselves can alleviate information asymmetries, and thus can form the basis of new forms of low cost lending. These tools together are enabling a new ecosystem of digital financial services.

# 7. REFERENCES

Arráiz, I., Bruhn, M., & Stucchi, R. (2017). Psychometrics as a Tool to Improve Credit Information. *The World Bank Economic Review*, *30*(Supplement_1), S67–S76. https://doi.org/10.1093/wber/lhw016

Berger, A. N., & Udell, G. F. (2006). A more complete conceptual framework for SME finance. *Journal of Banking & Finance*, *30*(11), 2945–2966.

Björkegren, D. (2010). "Big data" for development. Proceedings of the CEPR/AMID Summer School. Retrieved from http://dan.bjorkegren.com/files/CEPR_Bjorkegren.pdf

Björkegren, D., & Grissen, D. (2018). The Potential of Digital Credit to Bank the Poor. *American Economic Association Papers and Proceedings*.

Blumenstock, J., Cadamuro, G., & On, R. (2015). Predicting poverty and wealth from mobile phone metadata. *Science*, *350*(6264), 1073–1076.

Breiman, L. (2001). Random Forests. *Machine Learning*, *45*(1), 5–32. https://doi.org/10.1023/A:1010933404324

Breiman, L., & Cutler, A. (2006). *randomForest*. Retrieved from http://stat-www.berkeley.edu/users/breiman/RandomForests

Butler, D. (2013). When Google got flu wrong. *Nature News*, *494*(7436), 155. https://doi.org/10.1038/494155a

Carlson, S. (2017). Dynamic Incentives in Credit Markets: An Exploration of Repayment Decisions on Digital Credit in Africa. *Working Paper*.

Demirguc-Kunt, A., Klapper, L., Singer, D., & Van Oudheusden, P. (2014). The Global Findex Database. World Bank. Retrieved from http://www.worldbank.org/en/programs/globalfindex

Francis, E., Blumenstock, J., & Robinson, J. (2017). Digital Credit: A Snapshot of the Current Landscape and Open Research Questions. *CEGA White Paper*.

FSD. (2016). FinAccess Household Survey.

Gonzalez, M. C., Hidalgo, C. A., & Barabasi, A.-L. (2008). Understanding individual human mobility patterns. *Nature*, *453*(7196), 779–782. https://doi.org/10.1038/nature06958

Hand, D. J. (2009). Measuring classifier performance: a coherent alternative to the area under the ROC curve. *Machine Learning*, *77*(1), 103–123.

Isaacman, S., Becker, R., Caceres, R., Kobourov, S., Martonosi, M., Rowland, J., & Varshavsky, A. (2011). Identifying Important Places in People's Lives from Cellular Network Data. In K. Lyons, J. Hightower, & E. Huang (Eds.), *Pervasive Computing* (Vol. 6696, pp. 133–151). Springer. Retrieved from http://www.springerlink.com/content/r14x8r7573738143/abstract/

ITU. (2011). *World telecommunication/ICT indicators database*. International Telecommunication Union.

Lazer, D., Kennedy, R., King, G., & Vespignani, A. (2014). The Parable of Google Flu: Traps in Big Data Analysis. *Science*, *343*(6176), 1203–1205. https://doi.org/10.1126/science.1248506

Lu, X., Wetter, E., Bharti, N., Tatem, A. J., & Bengtsson, L. (2013). Approaching the Limit of Predictability in Human Mobility. *Scientific Reports*, *3*. https://doi.org/10.1038/srep02923

Ng, A. (2011). Machine Learning and AI via Brain simulations.

NPR. (2015). How Cellphone Use Can Help Determine A Person's Creditworthiness. *Morning Edition*. Retrieved from https://www.npr.org/2015/08/04/429219691/how-cellphone-usage-can-help-determine-a-person-s-credit-worthiness

Onnela, J.-P., Saramäki, J., Hyvönen, J., Szabó, G., Lazer, D., Kaski, K., … Barabási, A.-L. (2007). Structure and tie strengths in mobile communication networks. *Proceedings of the National Academy of Sciences*, *104*(18), 7332–7336. https://doi.org/10.1073/pnas.0610245104

Palla, G., Barabási, A.-L., & Vicsek, T. (2007). Quantifying social group evolution. *Nature*, *446*(7136), 664–667. https://doi.org/10.1038/nature05670

Pedro, J. S., Proserpio, D., & Oliver, N. (2015). MobiScore: Towards Universal Credit Scoring from Mobile Phone Data. In *User Modeling, Adaptation and Personalization* (pp. 195–207). Springer, Cham. https://doi.org/10.1007/978-3-319-20267-9_16

Soto, V., Frias-Martinez, V., Virseda, J., & Frias-Martinez, E. (2011). Prediction of Socioeconomic Levels Using Cell Phone Records. In J. A. Konstan, R. Conejo, J. L. Marzo, & N. Oliver (Eds.), *User Modeling, Adaption and Personalization* (pp. 377–388). Springer Berlin Heidelberg. https://doi.org/10.1007/978-3-642-22362-4_35

Suri, T., Jack, W., & Stoker, T. M. (2012). Documenting the birth of a financial economy. *Proceedings of the National Academy of Sciences*, *109*(26), 10257–10262. https://doi.org/10.1073/pnas.1115843109

Totolo, E. (2018). The Digital Credit Revolution in Kenya: An Assessment of Market Demand, 5 Years On. CGAP. Retrieved from http://www.findevgateway.org/library/digital-credit-revolution-kenya-assessment-market-demand-5-years

**Figure 1: Default Rate by Proportion of Borrowers Accepted**

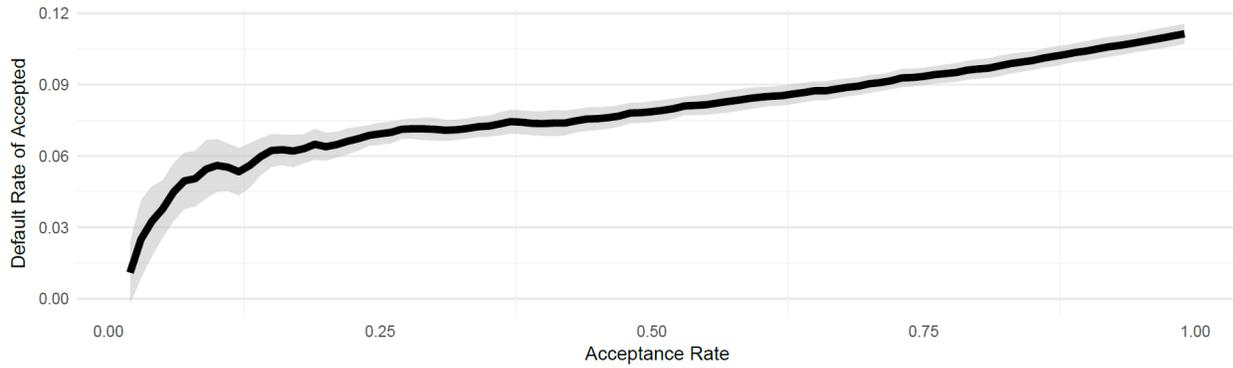

Phone indicators using the conservative random forest weekly ensemble model (CDR-W). Line shows mean, and ribbon standard deviation, of results from multiple fold draws.
Source: authors' analysis from telecom data.

**Figure 2: Receiver Operating Characteristic Curve**

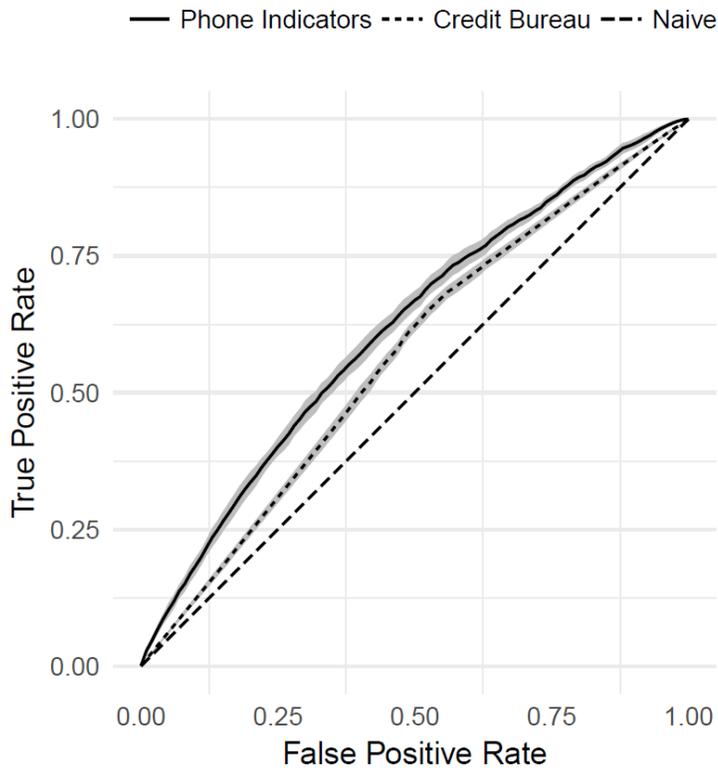

Credit bureau uses the highest performing stepwise logistic model. Phone indicators use the conservative random forest weekly ensemble model (CDR-W). Line shows mean, and ribbon standard deviation, of results from multiple fold draws.

Source: authors' analysis from telecom data.

# Figure 3: Performance by Level of Formalization

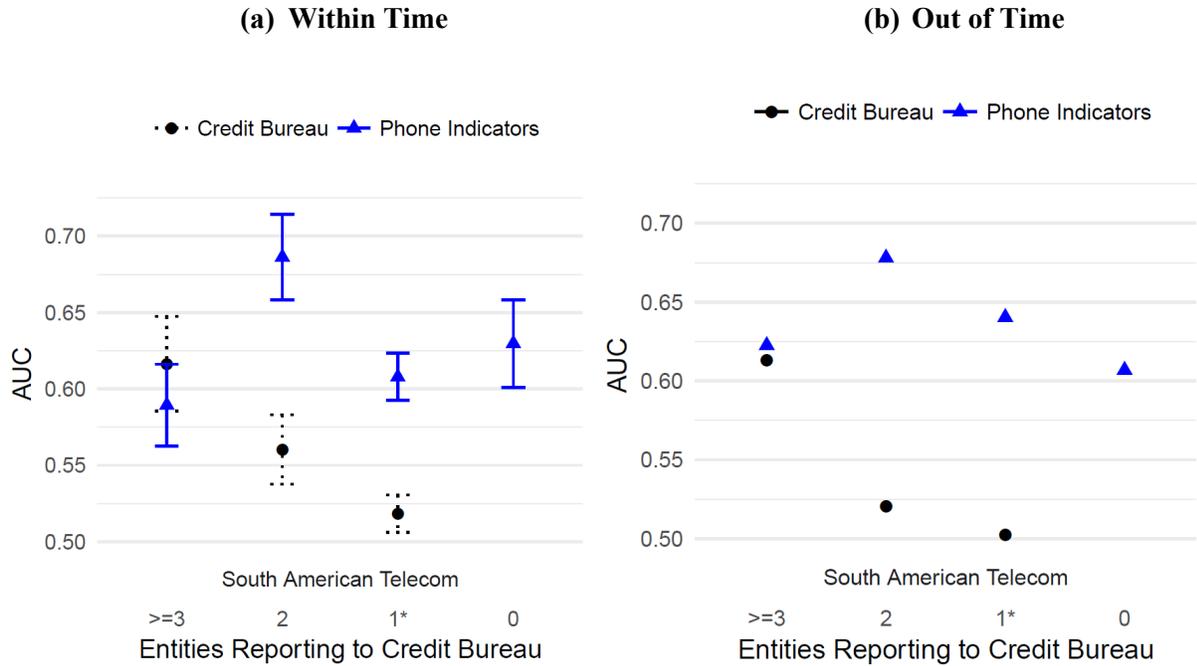

1*: either one entity reporting, or has a file at the credit bureau which may include previous activity but zero entities are currently reporting. Comparison of the highest performing bureau model (logistic stepwise) and most conservative phone indicator model (CDR-W Random Forest). Models are trained on all individuals but the omitted fold and area under the receiver operating characteristic curve (AUC) is reported for the subset of individuals within the omitted fold with the given number of entities reporting to the credit bureau. For out of sample estimates, the point shows the mean, and error bars standard deviation, of results from multiple fold draws. Out of time estimates use the offset indicators and only have a single fold draw.
Source: authors' analysis from telecom data.

## Table 1: Description of Individuals

|  | Mean | SD | Median |
|---|---|---|---|
| **Country GDP per capita ($, approx.)** | 6,000 | | |
| **Borrowers** | | | |
| Gender is female (%) | 39 | - | - |
| Age (years) | 35.8 | 12.8 | 34.0 |
| **Has a mobile phone (%)** | 100 | - | - |
| **Credit bureau record (%)** | 85 | - | - |
| Entities reporting: | | | |
| At least one (%) | 59 | | |
| At least two (%) | 31 | | |
| At least three (%) | 16 | | |
| **Average weekly mobile phone use** | | | |
| Calls out, number | 32.0 | 25.6 | 26.0 |
| Calls out, minutes | 41.6 | 39.9 | 32.0 |
| SMS sent | 31.3 | 26.3 | 24.4 |
| Days of mobile phone data preceding plan switch | 107 | 14 | 112 |
| **Credit** | | | |
| Default Rate (%) | 11 | - | - |
| N | 7,068 | | |

Source: authors' analysis from telecom data.

## Table 2: Individual Features

|  | Correlation with repayment | t-stat | Number of Features |
|---|---|---|---|
| **Demographics** |  |  | 2 |
| Age | 0.073 | 2.35 |  |
| Female | -0.039 | -1.26 |  |
| **Credit Bureau** |  |  | 36 |
| Has a credit bureau record | -0.022 | -1.89 |  |
| Summary score (lower is better) | -0.072 | -6.15 |  |
| Fraction of debt lost | -0.046 | -3.86 |  |
| **Phone usage** |  |  | 5,541 |
| *Categories* | *High performing example feature:* |  |  |
| Periodicity | -0.163 | -5.27 | 796 |
| | SMS by day, ratio of magnitudes of first fundamental frequency to all others | | |
| Slope | 0.126 | 4.06 | 44 |
| | Slope of daily calls out | | |
| Correlation | 0.111 | 3.57 | 224 |
| | Correlation in SMS two months ago and duration today | | |
| Variance | -0.104 | -3.34 | 4,005 |
| | Difference between 80$^{th}$ and 50$^{th}$ quantile of SMS use on days SMS is used | | |
| Other | 0.100 | 3.07 | 542 |
| | Number of important geographical location clusters | | |

Source: authors' analysis from telecom data.

# Table 3: Model Performance

| Dataset: | Main Results | | | Check |
|---|---|---|---|---|
| | Standard Indicators | | | Offset Indicators |
| Performance: | Out of Sample (5 fold CV) | | | Out of Time (train early period, test late) |
| Sample: | All | Has Bureau Records | No Bureau Records | All |
| | AUC | AUC | AUC | AUC |
| **Baseline Model** | | | | |
| **Credit Bureau** | | | | |
| Random Forest | 0.516 | 0.509 | - | 0.507 |
| Logistic, stepwise BIC | 0.565 | 0.565 | - | 0.550 |
| **Our Models** | | | | |
| **Phone indicators (CDR)** | | | | |
| Random Forest | 0.710 | 0.708 | 0.719 | 0.631 |
| Logistic, stepwise BIC | 0.760 | 0.759 | 0.766 | 0.595 |
| **Phone indicators, within-week variation (CDR-W)** | | | | |
| Random Forest Weekly Ensemble | 0.616 | 0.614 | 0.630 | 0.641 |
| OLS FE, stepwise BIC | 0.633 | 0.634 | 0.631 | 0.593 |
| **Combined** | | | | |
| **Credit bureau and phone indicators** | | | | |
| Random Forest | 0.711 | 0.708 | - | 0.642 |
| Logistic, stepwise BIC | 0.772 | 0.770 | - | 0.616 |
| **Credit bureau and phone indicators, within-week variation** | | | | |
| Random Forest Weekly Ensemble | 0.618 | 0.616 | - | 0.639 |
| OLS FE, stepwise BIC | 0.645 | 0.645 | - | 0.586 |
| Default Rate (%) | 11 | 12 | 10 | |
| N | 7,068 | 6,043 | 1,025 | 6,975 |

Standard indicators evaluate out of sample performance using 5-fold cross validation, averaged over fold draws. Offset indicators are derived from only half of the data (the first half for early transitions; the last half for late transitions); the out of time model is estimated on the early half of transitions and tested on the late half. AUC represents the area under the receiver operating characteristic curve, and BIC the Bayesian Information Criterion. For middle two columns, model is trained on all individuals except the omitted fold, and performance is reported for the given subsample within the omitted fold. Source: authors' analysis from telecom data.